\documentclass[dvips,12pt,aps,nofootinbib,preprint]{revtex4-1}
\usepackage{amsmath,amssymb,epsf,epsfig,amsthm,bm,graphicx,subfigure}
\usepackage[hmargin=2.15cm,vmargin=2.15cm]{geometry}

\theoremstyle{theorem}
\newtheorem{theorem}{Theorem}[section]

\newtheorem{corollary}[theorem]{Corollary}

\theoremstyle{definition}

\newtheorem{definition}[theorem]{Definition}

\newtheorem{remark}[theorem]{Remark}

\newcommand{\bX}{\bm{X}}

\newcommand{\bU}{\bm{U}}

\newcommand{\bhX}{\hat{\bX}}
\newcommand{\fl}{\mathbb{E}}

\newcommand{\peu}{{{}^\perp\! U}}
\newcommand{\pau}{{{}^\|\!U}}

\def\bi{\begin{itemize}}
\def\ei{\end{itemize}}

\begin{document}

\author{Steven Willison}

\address{CENTRA, Departamento de F\'{i}sica,
Instituto Superior T\'{e}cnico - IST, Universidade T\'{e}cnica de
Lisboa - UTL, Av. Rovisco Pais 1, 1049-001 Lisboa, Portugal}

\email{steven.willison@ist.utl.pt}

\title{A Re-examination of the isometric embedding approach to
General Relativity}
\begin{abstract}
We consider gravitational field equations which are Einstein
equations written in terms of embedding coordinates in some higher
dimensional Minkowski space. Our main focus is to address some
tricky issues relating to the Cauchy problem and possible
non-equivalence with the intrinsic Einstein theory. The well known
theory introduced by Regge and Teitelboim in 9+1 dimensions is
cast in Cauchy-Kowalevskaya form and therefore local existence and
uniqueness results follow for \emph{analytic} initial data. In
seeking a weakening of the regularity conditions for initial data,
we are led naturally to propose a 13+1 dimensional theory. By
imposing an appropriate conserved initial value constraint we are
able, in the neighbourhood of a generic (free) embedding, to
obtain a system of nonlinear hyperbolic differential equations.
The questions of long time or global existence and uniqueness are
formidable, but we offer arguments to suggest that the situation
is not hopeless if the theory is modified in an appropriate way.
We also  present a modification of the perturbation method of
G\"{u}nther to weighted Sobolev spaces, appropriate to noncompact
initial data surfaces with asymptotic fall-off conditions.
\end{abstract}

\maketitle
\section{Introduction}

The isometric embedding of a riemannian or pseudo-riemannian
manifold into an ambient flat space is a mathematical subject much
explored over the last centuary. It also crops up occasionally in
various guises in the context of gravitational physics. Manifolds
were originally concieved of as embedded surfaces but after
seminal work of the nineteenth centuary geometers the intrinsic
viewpoint emerged. Full equivalence between the two viewpoints was
established by Whitney in the case of abstract differentiable
manifolds and by Nash \cite{Nash56} for manifolds with a
Riemannian metric. The latter equivalence has been extended to
pseudo-riemannian manifolds by various people
\cite{Gromov_Rokhlin,Greene70,Clarke70,Mueller:2008gc}. Therefore,
in principle, any physics which can be formulated in terms of
curved manifolds, can alternatively be reformulated in terms of
curved submanifolds of flat space(-time) and vice versa.

In Ref. \cite{ReggeTeitelboim75} Regge and Teitelboim considered
the Hamiltonian formulation of General Relativity (GR), observing
a similarity between the constraint algebra of GR and that of the
bosonic string. (In strings, the auxiliary flat space-time
structure plays a key role in determining a clever choice of
non-local operators.) This motivated a study of embedded GR as a
possible route to canonical quantisation via the Dirac procedure.
The constraint algebra has been subsequently worked out in detail
\cite{PastonFranke07,PastonSemenova10} but the question of
quantisation remains unclear. Similarly in Ref. \cite{Faddeev09} a
10-vector formulation was proposed, effectively introducing 10
vielbeins to describe four dimensional space-time, leading to a
system very similar to the embedded one. This was motivated by
analogy with the Weinberg-Salaam theory of massive vector field,
which can be cured of  non-renormalisability by the Higgs
mechanism. Generally, we may say that the currently well
understood approaches to quantisation make some use of the
properties of Minkowski space and particularly the Lorentz group.
As such one hopes that by writing gravity as some kind of field
theory living in Minkowski space, the quantum theory may be made
more tractable. Some works along these lines are Refs.
\cite{Pavsic-1986},\cite{Paston:2011db}. The possibility to relate
the Lorentz symmetry of the embedding space with the internal
symmetries of particle physics was explored in refs
\cite{Joseph:1962zz}.

Perhaps a less ambitious hope is that the embedding point of view
may provide new methods to prove results in mathematical
relativity.
It may also be that an embedded theory
can shed light on certain physical notions which do not sit well
with the intrinsic tensorial formulation of gravity, such as a
local notion of gravitational energy or the nonlinear completion
of massive gravity.

In this article we will not venture into such possible
applications, but will discuss some basic problems of existence
and uniqueness which potentially plague any embedded theory of
gravity. These kind of problems were known from the beginning, see
e.g. Refs \cite{ReggeTeitelboim75},\cite{DeserPiraniRobinson76}.
By reinvestigating the mathematical foundations of the theory in
some detail we are able to clarify to some extent the situation.
The point of view adopted here has some similarities with that of
Ref. \cite{Bustamante05} in that we emphasise the so called free
embedding as the central idea. Our concern here is with the
physical theory, but we find it worthwhile to get a little into
the nuts and bolts of the various embedding theorems along the
way. Since this is rather a large and complicated subject, and
quite removed from standard methods in Relativity, we shall aim to
consider in detail only those ideas which we will need, and to
introduce them in a pedagogical way. For more detailed reviews see
Refs. \cite{Friedman_Review,Gromov_Rokhlin,Andrews,HanHong06}.


\subsection{Einstein-Hilbert action and embedding variables}

We shall consider an $n$-dimensional Lorentzian space-time
$(\mathcal{M}^n, g_{\mu\nu})$ and embeddings of (subsets of)
$\mathcal{M}$ into Minkowski space $\mathbb{E}^{N-1,1}$. If at
some point $p\in M$ there exists a neighbourhood ${\cal U}\subset
\mathcal{M}$ of $p$ with local coordinates $(\xi^\mu)$ and a map
$X: {\cal U} \to \mathbb{E}^{N-1,1}$ such that
\begin{gather}\label{gequalsdX}
\xi^\mu \to X^A(\xi)\, ,\qquad \eta_{AB} \frac{\partial
X^A}{\partial \xi^\mu}\frac{\partial X^B}{\partial \xi^\nu} =
g_{\mu\nu}\, ,
\end{gather}
and $X({\cal U})$ is a submanifold,
then we say that $M$ admits a local isometric embedding $X$ at
$p$. Naturally the case of most interest in physics is $n =4$. The
choice of $N$ is more a matter of circumspection. For a given
metric, we will have 10 equations, 4 of which may be considered
redundant due to coordinate transformations. Since transformations
$X\to X^\prime$ which map the sub-manifold $X({\cal U})$ to itself
are also redundant, giving also 4 redundant degrees of freedom, we
might expect $N\geq 10$ is necessary to avoid fatally
over-determining the equations. The same counting of degrees of
freedom applies if, instead of embedding a fixed metric, we are
solving Einstein's equations for $X$. More careful arguments for
choice of $N$ can be made and will be given in what follows.

Following Ref. \cite{ReggeTeitelboim75} we can attempt to describe
gravity with $X(\xi)$ as the fundamental variable instead of a
fundamental metric field\footnote{Whenever it is necessary to make
the distinction clear, we will use $g$ to denote an induced
metric, which is implicitly a function of $\partial X$ via
(\ref{gequalsdX}) and ${\frak g}$ to denote a fundamental metric
field. Likewise we distinguish between curvatures $R_{\mu\nu\kappa
\lambda}$ and ${\cal R}_{\mu\nu\kappa \lambda}$ and Einstein
tensors $G_{\mu\nu}$ and ${\cal G}_{\mu\nu}$.} ${\frak g}$.
Beginning with the Einstein-Hilbert action
\begin{gather}
 I[X,\phi_\text{matter}] = \int \sqrt{-{ g} }\,  {R}  \, d^4\xi
 + \int {\cal L}_\text{matter}\, ,
\end{gather}
and assuming that the matter Lagrangian depends on $X$ only
implicitly through $g_{\mu\nu}$, the Euler Lagrange equation
$\delta I/\delta X^A =0$ gives
\begin{gather}\label{contracted_Einsten}
 X^A_{;\mu\nu} ( G^{\mu\nu} -  T^{\mu\nu}) = 0\, .
\end{gather}
There are three issues preventing us from immediately concluding
that this is equivalent to the metric formulation of General Relativity: \\
\textbf{Issue i)} If the second fundamental form $X^A_{;\mu\nu}$
is not invertible in the appropriate sense, there are more
solutions besides $G^{\mu\nu}  =
T^{\mu\nu}$; \\
Even if we suppose $G^{\mu\nu}  = T^{\mu\nu}$ holds:
\\
\textbf{Issue ii)} Degrees of freedom may be missing, due to the
impossibility of embedding certain spacetimes.\\
\textbf{Issue iii)} There may be extra degrees of freedom due to
the possibility of isometric bending\footnote{Suppose $X_t$ is a
one-parameter family of small deformations of $X$ such that $X_0 =
X$ and  $X_t(M) \neq X(M)$ for $t>0$ and the induced metrics are
$g_t$ and $g$ respectively. If $(\mathcal{M},g)$ is isometric to
$(\mathcal{M},g_t)$ for some interval $[0,a]$ of $t$ we shall say
that $X_t$ is an isometric bending of $X$.}.

Due to well known theorems, issue ii) can be resolved by choosing
$N$ large enough. For local embeddings $N = 10$ is sufficient for
analytic metrics or $N = 14$ for smooth metrics.

Issue iii) is really two issues: firstly there may be extra
physical degrees of freedom, which is not necessarily undesirable;
secondly there may be nonphysical degrees of freedom with ill
defined evolution. To illustrate what can go wrong consider the
following: let $N=5$ and choose for initial data the surface $X^0
= X^5 =0$, and a time-like normal vector field along which we are
to develop our space-time, which we shall take to be
$\partial/\partial X^0$. Thus we specify the intrinsic metric to
be flat and extrinsic curvature to be zero. Let $T^{\mu\nu}=0$.
From the point of view of intrinsic geometry, the unique maximal
development of this data is Minkowski space $\mathbb{E}^{3,1}$.
From the embedding point of view however there is not uniqueness.
One possible development is the hypersurface $X^5 = 0$, but any
hypersurface $X^5 = f(X^0)$ satisfying $f^\prime(0) = 0$,and
$|f^\prime| \leq C$ with $C<1$ will also be a solution. Any
complete surface of this form is globally isometric to
$\mathbb{E}^{3,1}$. One might suspect from this example that
nonuniqueness is generic and that it becomes worse in higher
dimensions. But the reality turns out to be quite the contrary.
The nonuniqueness here is in fact a consequence of the
special way in which the initial surface is embedded, which is unavoidable in low dimensions but becomes \emph{non-generic} in higher dimensions.
If we choose $N = 10$ then we find that a generic choice of (analytic)
embedding and normal vector field
constitutes initial data for a well defined local evolution
problem. This shall be shown in detail in section
\ref{analytic_section}. For higher $N$, nonuniqueness will
reemerge, but there is a possibility to eliminate it for $N =14$
with a judiciously imposed constraint, which will be discussed in
section \ref{13+1_section}.

To guarantee the existence of global embeddings in principle
requires larger $N$, in which case problems of non-uniqueness may
become formidable. We pick up on this matter briefly in section
\ref{Large_N_section}.

The type of embedding mentioned above is basically what is called
a free embedding, defined in section \ref{Free_section}. If $N=
10$ the space-like sections will generically be freely embedded. It
is well known that, if we are in an open set of the function space
where this property holds, this partially resolves issue i),
subject to imposing the constraints $G_{0\mu}=T_{0\mu}$ by hand.
If $N\geq 14$ the space-time is generically freely embedded and
issue i) is resolved if we assume we are in a neigbourhood of such
a generic solution.
However there is a subtlety due to the fact that genericity in the sense of space-time free condition
is not a condition purely on initial data, since it depends on
second time derivatives of the embedding\cite{Private_comm_Paston}. We return to this point in  section \ref{13+1_section}
where we discuss the 14 dimensional theory, which is of necessity a background dependent theory.

\subsection{Intrinsic and extrinsic geometry}

\textbf{Notation:} For brevity, we shall use a vector space
notation both for position vectors in $\mathbb{E}^{N-1,1}$ and
tangent vectors, without distinction. Let $p$ be a point in
$\mathcal{M}$ with coordinates $(\xi^\mu)$. Its image, $X(p)\in
\mathbb{E}$ , is represented by a position vector $\bX = (X^A)$.
The tangent space $T_{X(p)}\mathbb{E}$ is decomposed into
components tangential and orthogonal to $X(\mathcal{M})$:
$T_{X} (\fl) = T_{X}^{\|}\oplus T_{X}^{\perp}$.
 The vectors
$\bX_{,\mu}$ form a basis of $T_{X}^{\|}$. The dot product
$\bm{V}\cdot \bm{W}$ denotes $\eta_{AB}V^AW^B$.

Basic geometrical quantities are the metric $g_{\mu\nu} =
\bX_{,\mu}\cdot \bX_{,\nu}$, Christoffel symbol (of the first
kind) $\Gamma_{\mu\, \rho\sigma} =
  \bX_{,\mu}\cdot \bX_{,\rho\sigma}$
and second fundamental form $\bX_{;\mu \nu}$. Here $;$ denotes
covariant differentiation with respect to $g$. The Riemann
curvature tensor is obtained from the Gauss formula:
\begin{gather}
 R_{\mu\nu\rho\sigma} = \bX_{;\mu\rho} \cdot \bX_{;\nu\sigma} -  \bX_{;\mu\sigma} \cdot \bX_{;\nu\rho}
\end{gather}
It is important to keep in mind that the second fundamental form
is normal-vector valued:
\begin{gather*}
 \bX_{;\mu\nu}\cdot \bX_{,\sigma} = 0\, .
\end{gather*}
In fact $\bX_{;\mu\nu}$ is the projection of $\bX_{,\mu\nu}$ onto
$T_{X}^{\perp}$.

\subsection{Free embeddings}\label{Free_section}

Let $M$ be of dimension $n$ and $\mathbb{E}$ be some flat space of
dimension $N$ and unspecified signature. Consider a given
embedding $\hat{X}:\mathcal{M}\to \mathbb{E}$ and some perturbed
embedding $X$ such that $\bX = \bhX + \bU$. The induced metric
will be perturbed $g_{\mu\nu} = \hat{g}_{\mu\nu}+h_{\mu\nu}$, with
\begin{align}
 h_{\mu\nu} & = 2\bhX_{,(\mu}\cdot \bU_{,\nu)} + \bU_{,\mu}\cdot
 \bU_{,\nu}   \nonumber
 \\
 & = 2(\bhX_{,(\mu}\cdot \bU)_{|\nu)} - 2 \bhX_{|\mu\nu}\cdot \bU + \bU_{,\mu}\cdot
 \bU_{,\nu}\label{starting_point}
\end{align}
where here "$|$'' denotes the covariant derivative using some
arbitrary background, not necessarily the physical background
$\hat{g}$. In order to motivate the free embedding, we neglect for now the nonlinear part, considering the
following system for $\bU$:
\begin{align*}
 \bhX_{,\mu}\cdot \bU & = 0\, ,
\\
 \bhX_{|\mu\nu}\cdot \bU & = -\frac{1}{2}h_{\mu\nu}\, .
\end{align*}
At the linearised level the tangential components of $\bU$
generate automorphisms of $X(M)$ and therefore the first condition
can be regarded as a gauge fixing. Representing $(\hat{X}^A_\mu$,
$\hat{X}^A_{|(\mu\nu)})$ by $\hat{X}^A_\Lambda$ with index
$\Lambda = 1,\dots \frac{1}{2}n(n+3)$, this can be written in
matrix form:
\begin{align*}
 U^A \eta_{AB} \hat{X}^B_{\ \Lambda}  = H_\Lambda\, ,\qquad H_\Lambda :=
 (0,-h_{\mu\nu}/2)
\end{align*}
For $N\geq \frac{1}{2}n(n+3)$ we find that the linear system can
be solved for $\bU$ in terms of $h_{\mu\nu}$ provided that
$\det{(\bhX_\Omega\cdot \bhX_\Lambda)}\neq 0$. If furthermore
$N=\frac{1}{2}n(n+3)$ then the solution is unique: $\hat{X}^A_{\
B} := (\hat{X}^{A}_\mu$, $\hat{X}^{B}_{|(\mu\nu)})$ is an
invertible square matrix and the solution is $U^A =
(\hat{X}^{-1})^A_{\ B} H^B$.


\begin{definition}
An embedding $X:{\cal U}\to \mathbb{E}$ is said to be free at $p$
if $\det{(\bX_\Omega\cdot \bX_\Lambda)}\neq 0$. If the condition
holds for all $p\in {\cal U}$ then $X$ is a \emph{free embedding}.
\end{definition}

For brevity, a free isometric embedding will be denoted FIE.

\begin{definition}
 For $p\in {\cal U}$ and an embedding $X:{\cal U}\to \mathbb{E}$, the vector
 subspace $\mathcal{O}_{X(p)} : = \text{span} (\bX_{,\mu} , \bX_{,\rho \sigma}) \in T_{X(p)}\mathbb{E}$
 is called the \emph{osculating space}.
\end{definition}

\begin{remark}\label{free_remark}
 An equivalent definition of free embedding is
 that: i) $\mathcal{O}_{X(p)}$ is not a null surface of
 $T_{X(p)}\mathbb{E}$ and; ii)
  the $(\bX_{\mu}, \bX_{,\mu\nu})$ are a set of linearly independent
 vectors, i.e. $\dim \mathcal{O}_{X(p)} = n(n+3)/2$.
 For an embedding of a Lorentzian manifold into $\mathbb{E}^{N-1,1}$,
 condition i) is automatic. But in the next section we will consider embedding of
 a Riemannian initial value surface into $\mathbb{E}^{N-1,1}$, where condition i) is nontrivial.
\end{remark}

\begin{remark}
A free embedding requires $N\geq n(n+3)/2$ dimensions. If $X$ is
free and $N=n(n+3)/2$, then $(\bX_{\mu}, \bX_{,\mu\nu})$ form a
basis of $T_{X(p)}\mathbb{E}$.
\end{remark}

We can now see why free embeddings are relevant to issue
ii) of section A. In 13+1 dimensions, the perturbations $U$
about a free embedding are in 1-to-1 correspondence with
$h_{\mu\nu}$ at the linearised level. This is also true for finite
perturbations if they are suitably small \cite{Clarke70}.
Therefore there is the possibility to establish an equivalence
between the degrees of freedom by gauge fixing the isometric
bending in the same way one gauge fixes $h$ in the intrinsic
theory. Although it is not completely
satisfactory to restrict the function space to free embeddings,
they are generic in a certain sense. Point-wise, a set of
$n(n+3)/2$ vectors in $N\geq n(n+3)/2 $ dimensions will
generically be linearly independent. This remains true, in the sense of open sets on the function space, for vector
fields if they are continuous and bounded. We may therefore adopt
a background field approach, fixing a background embedding and
studying close-by solutions. The existence of an appropriate
background is guaranteed by the theorem:
\begin{theorem}[Greene\cite{Greene70}]\label{GreeneTheorem}
 Any Lorentzian  manifold $(\mathcal{M}_4,g)$ with $g\in C^m$,
 $m\geq 3$ or $C^\infty$ admits local free isometric
 embedding in $\mathbb{E}^{13,1}$.
\end{theorem}

The following weaker definition is also of some importance:
\begin{definition} Let ${\cal U}$ be
 Lorentzian and $X:{\cal U} \to \mathbb{E}^{N-1,1}$ be an embedding.
 If there exists a foliation of ${\cal U}$  by space-like surfaces with adapted
 frame $(e_a, e_0)$ of $T{\cal U}$ such that
 the vectors $(\bX_{,a}, \bX_{,0} ,\bX_{;ab})$ are linearly independent
 and span a non-null vector subspace
 of $T\mathbb{E}$, then we say that $X$ is \emph{spatially free}.
\end{definition}
A spatially free embedding requires $N\geq n(n+1)/2$. The
terminology is our own, but the idea is not new. This condition,
or rather the analogous one for Riemannian geometry
\cite{HanHong06,Jacobowitz74}, is important in establishing the
Janet-Cartan-Burstin theorem. In the context of the
Regge-Teitelboim theory, it has been used widely in the literature
and its importance explicitly emphasised in Ref.
\cite{Bustamante05}.

The Lorentzian version of the Janet-Cartan-Burstin theorem is the
following:
\begin{theorem}[Friedman\cite{Friedmann61}]
\label{FriedmanTheorem} Any Lorentzian manifold
$(\mathcal{M}_4,g)$ with $g$ analytic, admits an analytic local
isometric embedding in $\mathbb{E}^{9,1}$.\end{theorem}

Theorems \ref{GreeneTheorem} and \ref{FriedmanTheorem} are special
cases ($n =4$) of the original theorems which hold for any $n$ and
for more general pseudo-riemannian signature. For smooth metrics,
Theorem \ref{GreeneTheorem} is the best we can do. It is not known
whether the dimension can be reduced by dropping the requirement
that the embedding be free. In particular, we wish to stress that
a smooth version of Theorem \ref{FriedmanTheorem} is \emph{not}
available.

\section{Analytic evolution problem for the 9+1 dimensional theory}%
\label{analytic_section} 

The problem of embedding a given space-time into
$\mathbb{E}^{9,1}$ can be formulated as a Cauchy-Kowalewskaya
system. Likewise for the evolution of the metric under Einstein's
equations. Therefore it stands to reason that Einstein's equations
expressed in embedding variables are also of that form. However,
we find it worthwhile to briefly go through the details, not least
because it brings out the role of what we call \emph{admissible}
embeddings of the initial data surface. Basically, for a given
choice of extrinsic curvature $K_{ab}$, not every free embedding
of the initial data surface allows us to extend the embedding into
a space-time neighbourhood. The condition of admissibility is a
Lorentzian analogue of the one discussed in Ref.
\cite{Jacobowitz74} but the signature change makes for a quite
different conclusion- certain embeddings, i.e. those for which the
osculating space is space-like, are admissible for \emph{any}
choice of $K_{ab}$.

The theory introduced in Ref. \cite{ReggeTeitelboim75} and
developed extensively in the literature describes (local) time
evolution of a 3-membrane embedded in $\mathbb{E}^{9,1}$. The
Euler Lagrange variation of the Einsten-Hilbert action gives.
\begin{gather}\label{ReggeTeitelboimEquations}
 {\bX}_{;ab} (G^{ab} - T^{ab})
 + 2{\bX}_{;0a}(G^{0a}-T^{0a}) + {\bX}_{;00}(G^{00}- T^{00}) = 0 \, .
\end{gather}
Since we have only 10 embedding dimensions the $X_{;\mu\nu}$ can
not be all linearly independent, so as things stand  we may not deduce the Einstein
equations from the action principle. But we may proceed as follows:

Let $\Sigma$ be an analytic 3-manifold and $Y:\Sigma \to
\mathbb{E}^{9,1}$ be an analytic free embedding which induces a
riemannian metric $\gamma$. Let $p$ be a point in $\Sigma$. Since
$Y$ is free at $p$, we can take as basis for the tangent space
$T_{Y(p)} \mathbb{E}$ the set of vectors\footnote{In this section
$;$ and $|$ denotes the 4 dimensional covariant derivative w.r.t
$g$ and 3-dimensional covariant derivative w.r.t. $\gamma$
respectively.} $(\bm{Y}_{|a}, \bm{Y}_{|ab}, \bm{E}_\perp)$, where
$\bm{E}_\perp$ is orthogonal to ${\cal O}_Y$. By remark
\ref{free_remark} it follows that $\bm{E}_\perp$ is not null and
the $\bm{Y}_{|ab}$ are linearly independent, which implies in turn
that $P_{abcd}: = \bm{Y}_{|ab}\cdot \bm{Y}_{|cd}$ possesses a
unique inverse $P^{abcd}$ such that $P_{abcd} P^{cdef} =
\frac{1}{2}(\delta^{a}_e \delta^{b}_{f} + \delta^{a}_f
\delta^{b}_{e})$.

We assume that $\Sigma$ is a submanifold of an analytic 4-manifold
$M$. For the metric theory, one gives as initial data $(\Sigma,
\gamma_{ab}, K_{ab})$, subject to the constraints $G_{\mu0}
=T_{\mu0}$. For the Regge-Teitelboim theory on the other hand, we
will need to specify $\bm{Y}$ and some unit time-like vector field
$\bm{V}$ over $\Sigma$, and solve the equations
to find an embedding $X:M
\to \mathbb{E}^{9,1}$ satisfying $\bm{X}|_\Sigma = \bm{Y}$ and
$\bm{X}_{;0}|_\Sigma = \bm{V}$.
Suppose we are given $Y$ and $V$. Then
\begin{gather}
 K_{{a}{b}} = \bm{V}\cdot \bm{Y}_{|{a}{b}}
\end{gather}
uniquely determines $K$. However, if we wish to prescribe $\gamma$
and $K$, and are given some isometric embedding $Y$ of $(\Sigma,
\gamma)$, we need to face the issue of whether or not an
appropriate vector field $V$ exists.
\begin{definition}
Given  analytic $(\Sigma,\gamma, K )$, a free analytic embedding
$Y$ is \emph{admissible} if $\gamma_{ab}:=
\bm{Y}_{|a}\cdot\bm{Y}_{|b}$ and $\exists$ an analytic vector
field $V \in T \mathbb{E}|_\Sigma$ which satisfies at each point
on $\Sigma$:
 \\
  i) $\bm{V} \cdot \bm{Y}_{|a} = 0$;
  \\
  ii) $\bm{V} \cdot \bm{Y}_{|ab} = K_{ab}$;
  \\
  iii) $\bm{V}\cdot \bm{V} = -1$;
  \\
   iv) $\bm{V}$ is linearly independent of ${\cal O}_Y$.
\end{definition}
Note that if such a $V$ exists, it is unique. Conditions i) and
ii) require that $\bm{V}$ is decomposed as: $ \bm{V} = P^{abcd}
K_{ab} \bm{Y}_{|cd} + V^\perp \bm{E}_\perp$. Condition iv)
requires that $V^{\perp}$ be nonvanishing. Its value will be
determined by solving condition iii), i.e. $ -1 = P^{abcd}
K_{ab}K_{cd} +  \bm{E}_\perp \cdot \bm{E}_\perp (V^\perp)^2$ for
$V^\perp$ if a solution exists. If the osculating space is
space-like (i.e. $\bm{E}_\perp \cdot \bm{E}_\perp = -1$) then there
is always a nonvanishing solution. If the osculating space is
space-timelike, a solution exists if $P^{abcd} K_{ab}K_{cd} \leq
-1$ on $\Sigma$ (in terms of a local condition: if at a point $p
\in \Sigma$ we have the strict inequality $P^{abcd} K_{ab}K_{cd} <
-1$ then we can find an analytic vector field $V_{\perp}$ so that
the embedding is admissible in some open neighbourhood of $p$.).
\begin{remark}\label{admissible_Remark}
 If $(\Sigma,\gamma) \to \mathbb{E}^{9}$ is free,
 then the embedding $Y$ induced by identifying $\mathbb{E}^{9}$
 with a constant time surface of $\mathbb{E}^{9,1}$ will also be free,
 with space-like osculating space. Therefore $Y$ will be admissible for
 any $K_{ab}$. Since $(\Sigma, \gamma)$ always admits local analytic FIE
 into $\mathbb{E}^9$ \cite{Friedmann61}, a local admissible
 embedding always exists.
\end{remark}

Returning to (\ref{ReggeTeitelboimEquations}), and working in an
adapted frame such that $e_0$ is orthogonal to $\Sigma$, we impose
$(G_{\mu0}-T_{\mu0})|_\Sigma =0$ by hand (but see
\cite{PastonFranke07} where a modified action principle
implementing this was proposed). Then
(\ref{ReggeTeitelboimEquations}) reduces to
\begin{gather}
 G_{\mu\nu}(\partial X, \partial \partial X)  =T_{\mu\nu}
 \, .\label{embedded_Einstein_implicit}
\end{gather}
Equation (\ref{embedded_Einstein_implicit}) can be put into
(second order) Cauchy-Kowalevskaya form by further specialising to
a Gaussian normal coordinate basis. The embedding equations $g =
{\frak g}$ become $\bX_{,a}\cdot \bX_{,b} = {\frak g}_{ab}$,
$\bX_{,a}\cdot\bX_{,0} = {\frak g}_{a0} =0$ and $\bX_{,0}\cdot
\bX_{,0} ={\frak g}_{00} =-1$. Following a standard procedure
\cite{HanHong06,Jacobowitz74}, we can differentiate to obtain the
equivalent second order system:
\begin{gather}
 \bX_{,00}\cdot \bX_{,0} =0 \, , \quad \bX_{,00}\cdot \bX_{,a}
 =0\, ,\quad \bX_{,00}\cdot \bX_{,ab} = -\frac{1}{2} {\frak g}_{ab,00} +
 \bX_{,0a}\cdot\bX_{,0b}\, ,
\end{gather}
with initial value constraints
\begin{gather}\label{JanetCartanConstraints}
 \bX_{,a} \cdot \bX_{,b}= {\frak g}_{ab}\, ,\quad
 \bX_{,0}\cdot \bX_{,0} = -1 \, , \quad \bX_{,0}\cdot \bX_{,a} =0\,
 ,\quad \bX_{,0}\cdot \bX_{,ab}= -\frac{1}{2} {\frak g}_{ab,0}\, .
\end{gather}
If $X$ is spatially free then the matrix $M^A_B :=
(X^A_{,0},X^A_{,a}, X^A_{,(ab)})$ is invertible. If ${\frak g}$
were a given background field
then the system can immediately be put in Cauchy-Kowalevskaya
form. In order to deal with (\ref{embedded_Einstein_implicit}) we
have a bit more work to do, but it is also a standard
construction\cite{Choquet-Bruhat09}. In Gaussian normal
coordinates the Einstein equations take the form
\begin{align}
 \frac{1}{2} {\frak g}_{ab,00} = S_{ab} + {\cal F}_{ab}\, ,
 \\
 {\cal R}_{0\mu} = S_{0\mu}\, , \label{EinsteinConstraints}
\end{align}
where ${\cal F}_{ab}$ and ${\cal R}_{0\mu}$ are first order in
time derivatives and $S_{\mu\nu} -\frac{1}{2} {\frak g}_{\mu\nu}
S_{\sigma}^{\ \sigma}= T_{\mu\nu}$. (One can consider a combined
Einstein-matter system but for simplicity, the stress tensor will
be regarded as a given background field.) The combined system
\begin{gather}
 \bX_{,00}\cdot \bX_{,0} =0 \, , \quad \bX_{,00}\cdot \bX_{,a}
 =0\, ,\quad \bX_{,00}\cdot \bX_{,ab} = -S_{ab}-{\cal F}_{ab} +
 \bX_{,0a}\cdot\bX_{,0b}\, ,\label{combined1}
 \\
 \frac{1}{2} {\frak g}_{ab,00} = S_{ab} + {\cal F}_{ab}\, ,
\end{gather}
along with initial value constraints
(\ref{JanetCartanConstraints}) and (\ref{EinsteinConstraints}), is
equivalent to (\ref{embedded_Einstein_implicit}). We can write
(\ref{combined1}) as:
\begin{gather}
  X^A_{,00} = (M^{-1})^A_{B} F^B
\end{gather}
Where $F^B := (0,0, -S_{ab}-F_{ab} + \bX_{,0a}\cdot\bX_{,0b} )$
and $M^{-1}$ is an analytic function of $\bX_{,0},\bX_{,a}$ and
$\bX_{,ab}$.

\begin{theorem}\label{9+1CauchyProblemTheorem}
 Given analytic $(\Sigma, \gamma, K)$ satisfying the Hamiltonian
 and momentum constraints and an admissible embedding
 $Y:\Sigma \to \mathbb{E}^{9,1}$ then $\exists$ a
 neighbourhood ${\cal N}$ of $\Sigma$ in $\mathcal{M}$, homeomorphic to
 $\Sigma \times (-\tau, \tau)$, a
 nondegenerate Lorentzian metric $g_{\mu\nu}$ on ${\cal N}$ and an
 analytic isometric embedding $X:({\cal N},g)\to \mathbb{E}^{9,1}$
 such that:
$g_{\mu\nu}$ satisfies the Einstein equations
$G_{\mu\nu}=T_{\mu\nu}$ on ${\cal N}$;
  $X(\Sigma) = Y(\Sigma)$; the pullback of $g$ onto $\Sigma$
 is $\gamma$; the extrinsic curvature of $\Sigma$
 is equal to $K$.
 The solution $X$ is unique up to diffeomorphisms of
 ${\cal N}$ which reduce to the identity on $\Sigma$.
\end{theorem}

For any $p \in \Sigma$ we can introduce a gaussian normal
coordinate neighbourhood ${\cal U}$ of $p$ in $\mathcal{M}$. Since
$Y$ is admissible, we can find a unique $V$ such that
(\ref{JanetCartanConstraints}) is satisfied for $X(\Sigma)$ =
$Y(\Sigma)$ and $\bm{X}_{,0}|_\Sigma = \bm{V}$. The local result
follows by applying the Cauchy-Kowalevskaya theorem to the above
equation system. This implies local geometrical uniqueness up to
diffeomorphisms which reduce to the identity on ${\cal U}\cap
\Sigma$. Therefore the result can be be stated in the above
tensorial form, whereupon the global-in-space result follows in
standard fashion by considering an appropriate coordinate atlas on
$\Sigma$ and patching together solutions. $\Box$

Given $(\Sigma, \gamma, K)$, by remark \ref{admissible_Remark} we
can always find an admissible embedding in the neighbourhood of
any $p\in \Sigma$. However,without some assumptions on the
geometry and topology, a global admissible embedding may not
exist. This motivates the search for global admissible embeddings
${M }^3 \to \mathbb{E}^{9,1}$ where $M^3$ is a case of physical
interest. This is left as an open problem.

The fact that we obtain uniqueness does not mean that there is no
isometric bending. It means that there is no residual isometric
bending left over once we have fixed the initial surface. The
choice of $Y$ itself is not unique - from naive counting of
components we expect 4 degrees of isometric bending. The
appearance of extra independent initial data whose evolution is
well defined, suggests that we have more than the 2 physical
degrees of freedom. We could simply declare these extra degrees of
freedom to be pure gauge but it is not clear from our analysis why
this should be done. On the other hand it \emph{is} natural from
the Hamiltonian point of view - isometric bending is an invariance
of the action and so will not have a corresponding conjugate
canonical momentum. This manifests itself in terms of additional
constraints on the canonical variables. This approach is in fact
shown in \cite{PastonSemenova10} to be consistent and reproduce
the degrees of freedom of GR. In that reference, four (linear
combinations of) the constraints were isolated as generators of
isometric bending and were shown to form the ideal of the
constraint algebra.

It does not seem to be possible to relax the analyticity
assumption in the above results without some other modifications
to the theory. Even for $C^\infty$ initial data, no local
existence result is available. This is why we are reliant on the
Cauchy-Kowalevskaya theorem to establish existence and uniqueness.
This leads to what seems to be a fundamental theoretical weakness
of the 9+1 dimensional theory. This problem is very similar to one
raised in Ref. \cite{Anderson:2004zu} regarding the use of the
Campbell-Magaard theorem in physical applications. Basically, the
embedding theorem of Friedman, combined with the well-posedness of
the intrinsic Einstein equation system, is not strong enough to
deduce the well-posedness of the embedded Einstein system. This is
because the CK theorem does not guarantee continuous dependence on
the data. Furthermore, the domain of dependence property does not
strictly speaking make sense for analytic data. From the physical
point of view one might be content with an approximate local
domain of dependence result for analytic fields, whereby two
different sets of analytic data which agree closely in some small
enough region ${\cal U} \subset \Sigma$ will agree closely
regarding the developement $X({\cal D}^+({\cal U}))$. But in order
to be able to deduce this we would need continuous dependence.
Therefore it is unclear how to proceed with the 9+1 dimensional
theory.

Let us consider two possible approaches to smooth data:

1) A natural way to relax the regularity and attempt to obtain
continuous dependence is to formulate the problem in terms of a
(probably highly nonlinear) system of wave equations. This will
require an approach which is more democratic with respect to space
and time which ultimately requires the background to be free
rather than merely spatially free. For this it is necessary to go
to 13+1 dimensions.

2) At first glance one might think that Theorem
\ref{9+1CauchyProblemTheorem} can be strengthened to the smooth case by approximating the initial data by analytic functions, and then applying the implicit function theorem to obtain smooth solutions with the original data. The problem with this is that the implicit function theorem can only be invoked if the spacetime is freely embedded. If we follow the above construction which led to Theorem \ref{9+1CauchyProblemTheorem} in 13+1 dimensions, the analytic solution is not unique, and it turns out it can always be modified to a free embedding as described in Ref. Ref.\cite{Jacobowitz74}.

The two approaches are slightly different but complementary. The second enables us to immediately conclude local existence of solutions in 13+1 dimensions, but without giving any uniqueness result. Indeed there is extra gauge freedom to be fixed, as can be seen more clearly in the first approach. We will focus on the first approach, which will be elaborated on in section
\ref{13+1_section}. In any case, all roads seem to lead us to $N\geq 14$.

\section{The perturbation problem for isometric embeddings of
initial data surfaces}
\label{perturbation_section}

This section is somewhat of an aside, and can be omitted by the
reader wishing to proceed directly to the 13+1 dimensional theory.
It is included here since some of the ideas introduced provide
some context and inspiration for the methods used in the next
section.

A question of considerable interest regarding isometric embeddings
is the following: given $({\cal U},\hat{g})$ which admits a FIE
into $\mathbb{E}^N$, do all nearby metrics also admit such an
embedding? Well known proofs exist in which the concept of
``nearby" is defined in terms of $C^n$ or H\"{o}lder norm.
G\"{u}nther\cite{Gunther89} introduced a useful method to
effectively reduce the problem to an elliptic one. This method is
powerful because it can be used to solve compactly supported
metric perturbations in terms of compactly supported perturbations
of $X$. This in turn gave a reduction of the dimensions for Nash's
theorem\cite{Gunther90}. Our interests are somewhat different so
we will here give only a simplistic introduction which is
sufficient for our purposes. There are two quite different reasons
for our interest in this method, both of which arise when we
abandon the notion of analytic gravitational field. Firstly, it
becomes natural to formulate the field equations as a hyperbolic
system. In so doing, the method of G\"{u}nther provides a guide
for how to proceed by analogy. Our second reason relates to the
embedding of the initial data surface. We would like to be sure that all
close-by data can be achieved by perturbing it, so as to account
for all the degrees of freedom.

In Ref. \cite{Gunther89} Holder
spaces were used and by a trick the invertibility of the Laplacian minus a constant
on compact manifolds or with trivial boundary conditions was
exploited to obtain the perturbation result. However, for cases of physical
interest, such as asymptotically flat manifolds, we are interested
in special classes of non-compact manifolds with some fall-off
conditions at infinity.
So it would be desirable to have a perturbation result more
directly applicable to these manifolds by working with weighted
Sobolev spaces. As a preliminary step in this direction we
consider Riemannian 3-manifolds with trivial topology with
perturbations decaying appropriately at infinity\footnote{The
case of nontrivial topology should in principle be tractable, with
the appropriate generalisations of the weighted sobolev space and
(Lichnerowicz) laplacian.}.

A perturbation of an isometric embedding is described by
(\ref{starting_point}). Note that the equation remains valid if
$|$ is a covariant derivative with respect to any background
metric, not necessarily the metric we wish to perturb. The most
appropriate choice is the background metric $e$ used to define the
norm on the function space. Here we shall restrict ourselves to
manifolds with topology $\mathbb{R}^3$ so we take $e$ to be the
standard cartesian metric. Hence:
\begin{gather}\label{eucl_pert}
 h_{ab} = 2 (\bhX_{,(a}\cdot \bU)_{,b)} - 2 \bhX_{,ab} \cdot \bU
 + \bU_{,a}\cdot\bU_{,b}\, ,
\end{gather}
is valid globally.

In section \ref{Free_section} we saw that the perturbation result
at the linearised level follows immediately upon setting
$\bhX_{,\mu}\cdot\bU =0$. Now keeping the nonlinear terms, one
obtains
\begin{gather*}
 \bhX_{,a}\cdot \bU = 0\\
 \bhX_{,ab} \cdot \bU = -\frac{1}{2} h_{ab} +
 \bU_{,a}\cdot \bU_{,b}\, .
\end{gather*}
The existence of solutions for small $h$ and $U$ is not obvious,
since the nonlinear terms contain derivatives. The result can be
obtained by using a Nash-Moser type iteration scheme which
introduces smoothing operators to restore differentiability at
each step (see e.g. the appendix of ref \cite{Clarke70}). The
approach of Ref. \cite{Gunther89} dispenses with such
technicalities by introducing a more complicated gauge fixing of
$\bhX_{,\mu}\cdot\bU$ so as to obtain an elliptic system. We
introduce the Laplacian $\Delta = \delta^{ab}
\partial_a \partial_b$ and consider instead the system
\begin{align}
 \Delta(\bhX_{,a}\cdot \bU) & = - \Delta \bU\cdot\bU_{,a}\, ,
 \label{gunther_1}
\\
  \Delta (\bhX_{,ab} \cdot \bU ) & = -\frac{1}{2}\Delta h_{ab}
 + \bU_{,ac}\cdot
 \bU_{,b}^{\ \, c} - \bU_{,ab}\cdot \Delta \bU\, .
 \label{gunther_2}
\end{align}
By applying $\Delta$ to
(\ref{eucl_pert}) and imposing (\ref{gunther_1}) we obtain
(\ref{gunther_2}). Supposing that $\Delta^{-1}$ is well defined we
can write
\begin{gather}
 U = M^{-1} H + M^{-1} \Delta^{-1} Q(U)\, ,\label{gunther_matrix}
\end{gather}
where
\begin{gather*}
M = (\hat{X}^A_{,a} , \hat{X}^A_{,ab})\, ,\quad
 H = \left(%
\begin{array}{c}
  0 \\
  -\frac{1}{2} h_{ab}  \\
\end{array}\right)\, , \\ Q(U) = \left(%
\begin{array}{c}
  - \Delta \bU\cdot\bU_{,a} \\
  \bU_{,ac}\cdot
 \bU_{,b}^{\ \, c} - \bU_{,ab}\cdot \Delta \bU \\
\end{array}\right)\, .
\end{gather*}
If $N =9$ then $M^{-1}$ is unique. If $N>9$ we can choose the
unique $M^{-1}$ such that $M^{-1}H$ lies within the osculating
space ${\cal O}_{\hat{X}}$. A solution of (\ref{gunther_matrix})
will solve (\ref{eucl_pert}). The existence and uniqueness (modulo
the choice of $M^{-1}$) of such a solution can be established by
contraction mapping arguments.

Here we revisit the proof using weighted Sobolev spaces
$H_{p,\alpha}$. Notation and basic properties are in appendix
\ref{sobolev_appendix}. We will assume that an explicit FIE,
$\hat{X}$, of $(\mathbb{R}^3,\gamma)$ into $\mathbb{E}^N$ is known
which has reasonable asymptotic behaviour\footnote{ Here
reasonable means that $|K| + \|L\|_{q,\beta}$ is finite. We
include the constant term $K$, since this seems to be important in
finding examples, e.g. by some modification of the simple
Veronese type embedding $\xi \to X(\xi) =(\xi^1,\xi^2,\xi^3,
\frac{(\xi^1)^2}{\sqrt 2},\frac{(\xi^2)^2}{\sqrt 2},
\frac{(\xi^3)^2}{\sqrt 2}, \xi^1\xi^2,\xi^2\xi^3,\xi^2\xi^3)$.}. Then we obtain:

\begin{theorem}\label{Perturbation_Sobolev}
Let $\alpha,\beta >0$, $q\geq p\geq 4$. Given a FIE $\hat{X}$ for
($\mathbb{R}^3 $, $\gamma$)  such that $M^{-1} = K + L$ where $K$
is a constant matrix and $L \in H_{q,\beta}$, then there exists
some constant $C$ such that for any
 metric $\gamma+h$ such that $M^{-1}H \in
H_{p,\alpha}$ and satisfies
\begin{gather}\label{smallness}
  \| M^{-1}H\|_{p,\alpha}   (|K| + \|L\|_{q,\beta} ) \leq
  C\, ,
\end{gather}
 there exists a FIE $X$ for ($\mathbb{R}^3 $, $\gamma+h$) such that $\bX = \bhX +\bU$ with $U\in H_{p,\alpha}$ satisfying
\begin{gather}\label{smallness_U}
 \|U\|_{p,\alpha} \leq  \sqrt{\frac{\| M^{-1}H\|_{p,\alpha}}{|K| + \|L\|_{q,\beta} }
 }\, .
\end{gather}
\end{theorem}
With the above regularity assumptions the proof goes through much
the same as given in e.g. \cite{HanHong06},
and so we relegate it to appendix \ref{proof_appendix}. If we
further assume $h \in H_{p,\alpha}$ we obtain the following
corollary, which admits a more intuitive interpretation.
\begin{corollary}\label{Perturbation_Sobolev_corollary}
Under the additional hypothesis $h \in H_{p,q}$, Theorem
\ref{Perturbation_Sobolev}
 holds with (\ref{smallness}) and (\ref{smallness_U}) replaced by
\begin{gather}\label{smallness_h}
 \| h \|_{p,\alpha} \leq \frac{C}{(|K| + \|L\|_{q,\beta} )^2}
\end{gather}
and
\begin{gather}
 \|U\|_{p,\alpha} \leq D \sqrt{\| h\|_{p,\alpha} }
\end{gather}
respectively, for some constants $C$ and $D$.
\end{corollary}
Roughly speaking, $|K| + \| L\|_{q,\beta}$ is a measure of how
close the vectors $(\bhX_{,a},\bhX_{,ab})$ are to being linearly
dependent. So from (\ref{smallness_h}) we can say heuristically
that the ``more free" we make the background embedding, the larger
the perturbations for which we are able to exactly solve the
system.

It is important to note that we have only assumed some fall-off
conditions for the metric perturbation $h$, not for the background
$\gamma$. This includes as a special case all small perturbations within the
class $\gamma, h \in H_{p,\alpha}$, $p \geq 4$, $\alpha
>1/2$ of asymptotically flat metrics on $\mathbb{R}^3$ with well defined ADM total
energy. Such metrics are $C^2$ and
have an asymptotic fall-off faster than $r^{-\alpha}$.


\section{Proposal: 13+1 dimensional theory}\label{13+1_section}

Here we will argue that it is natural to generalise the
Regge-Teitelboim theory to 13+1 dimensions, being the minimal
dimension in which a local free embedding of space-time exists. If
we consider the theory in the neighbourhood of some freely
embedded space-time, there are two interesting features. The Euler-Lagrange
equations (\ref{contracted_Einsten}) are equivalent to the
Einstein equations in terms of the embedding coordinates. Secondly, with a bit of work, we can obtain
a system of nonlinear wave equations for the perturbed embedding coordinates $U$.
The first feature is subject to the following
caveat\cite{Private_comm_Paston}:  Given
initial data such that $X_{;a}, X_{;0}, X_{;ab},X_{;a0}$ are close to our background embedding,
we  we can not deduce a priori from (\ref{contracted_Einsten}) that $X_{;00}$ is
close to the background, and therefore we can not guarantee that $X$ is free.
We may impose the Hamiltonian constraint $G_{00} =
T_{00}$ by hand, whereupon (\ref{contracted_Einsten}) do reduce to
the Einstein equations. These are
linear in $X_{;00}$, but we can not solve algebraically for $X_{;00}$ because they only determine the projection of $X_{;00}$ onto span$(X_{;a}, X_{;0}, X_{;ab},X_{;a0})$ leaving one component undetermined. So the issue still remains that we have no a priori bound on $X_{;00}$. In view of this, we will defer the question of an action principle for the future, and simply look for a set of equations which will: reproduce the Einstein equations in some gauge fixing; completely determine the evolution of $X$; ensure that $X$ remains close to some freely embedded background, at least for small initial data and  for some finite time. Therefore we will necessarily have a background-dependent formulation. As a first step, we write  the Einstein equations as
$R_{\mu\nu} = S_{\mu\nu}$ i.e.
\begin{gather}\label{embedded_Einstein}
{\bm X}_{;\mu\nu} \cdot {\bm X} _{;\rho }^{\ \, \rho} -  {\bm
X}_{;\mu\rho} \cdot {\bm X} _{;\nu }^{\ \, \rho} =S_{\mu\nu}\, .
\end{gather}
This equation system does not belong to any of the well known
types and little can be said about it in its current form.

We shall expand the space-time metric about a background, which we
shall assume here to be flat, $g_{\mu\nu} = \eta_{\mu\nu} +
h_{\mu\nu}$. (If we take some other background the analysis is
basically the same in terms of obtaining a hyperbolic system but
the equations are considerably more complicated.) Let
$\hat{X}:({\cal U},\eta) \to \mathbb{E}^{13,1}$ be a FIE in a
neighbourhood ${\cal U}$ of $p \in \mathcal{M}$. The osculating
space spans the tangent space, with vectors
$\bhX_{,\mu}$ spanning $T_\|$ and vectors $\bhX_{;\mu\nu}$ spanning
$T_{\perp}$. A nearby solution $X$ will be a
FIE of $({\cal U},g)$ such that $X= \hat{X} +U$. Let us make use
of the notation:
\begin{gather}
 {}^\|\!U_\mu := \bhX_{,\mu}\cdot \bU\, , \qquad
 {}^\perp\!U_{\mu\nu} := \bhX_{,\mu\nu}\cdot \bU\, .
\end{gather}
With these definitions, $\bm{U}$ can be decomposed uniquely as:
\begin{gather}\label{U_decomposition}
 \bm{U} =  \peu_{\mu\nu} (P^{-1})^{\mu\nu\rho\sigma}
 \bhX_{,\rho\sigma} + \pau_{\mu} g^{\mu\nu} \bhX_{,\nu}\, ,
\end{gather}
where $P_{\mu\nu\rho\sigma} := \bhX_{\mu\nu}\cdot
\bhX_{\rho\sigma}$ and $(P^{-1})^{\mu\nu\rho\sigma} $ is the
unique inverse $(P^{-1})^{\mu\nu\rho\sigma} P_{\rho\sigma\kappa
\lambda} = \frac{1}{2} \left(\delta^\mu_\kappa \delta^\nu_\lambda
+ \delta^\nu_\kappa \delta^\mu_\lambda\right)$.

In the metric theory of GR, a standard
method\cite{Choquet-Bruhat09} to obtain a hyperbolic equation
system is to consider the reduced Einstein equations: ($\Gamma_\mu
:= g^{\rho\sigma} \Gamma_{\mu\, \rho\sigma }$)
\begin{gather}\label{CB_Reduced_Einstein}
 R_{\mu\nu} - \Gamma_{(\mu,\nu)} = S_{\mu\nu}
\end{gather}
along with initial value constraints
\begin{gather}
 \Gamma_\mu|_\Sigma = 0 \, \qquad G_{0\mu} |_\Sigma= 0\, .
\end{gather}
The constraints are conserved and therefore this system is
equivalent to the Einstein equations under the de Donder gauge
fixing of the coordinate freedom. The point being that $R_{\mu\nu}
= -\frac{1}{2}\tilde{\Box}g_{\mu\nu} + \Gamma_{(\mu,\nu)}+\cdots$,
where $\tilde{\Box} := g^{\mu\nu}\partial_{\mu}\partial_{\nu}$ and
the ellipsis means we are concerned only with the principal part.
Therefore (\ref{CB_Reduced_Einstein}) is hyperbolic about any
non-degenerate $g$.

Simply following the above approach will not quite serve for the
embedding theory. The problem can be seen from
(\ref{starting_point}) -  $h_{\mu\nu}$ is first derivative in $U$.
This means third derivatives of $U$ will show up in the reduced
Einstein equations. In order to get around this, we propose to
split the de Donder condition into two separate gauge fixing
conditions, something which  is possible precisely because we have
enlarged the embedding space to 14 dimensions.
We hope that what follows will convince the reader that this is
also a natural and useful thing to do.

Forgetting for a moment the Einstein equations, if we wished to
cast the embedding problem for a pre-prescribed metric $g = \eta +
h$ as a hyperbolic system we might perform a Lorentzian version of
(\ref{gunther_1}) and (\ref{gunther_2}), introducing the gauge
fixing $\Box{}^\|\!U_\mu = -{\bm U}_{,\mu}\cdot \Box {\bm U}$ so
as to obtain $-\frac{1}{2}\Box h_{\mu\nu} = \Box
{}^\perp\!U_{\mu\nu} + \cdots$. However, when we come to consider
the Einstein equations we wish instead to obtain
$-\frac{1}{2}\tilde\Box h_{\mu\nu} + \Gamma_{(\mu,\nu)} =
\tilde\Box {}^\perp\!U_{\mu\nu} + \cdots$. It turns out that the
relevant condition is: (this will be shown in detail below, see
(\ref{appropriate_form}))
\begin{gather*}
 \tilde\Box{}^\|\!U_\mu + {\bm U}_{,\mu}\cdot \tilde\Box
{\bm U} -\Gamma_\mu = g^{\rho\sigma}
\left({}^\perp\!U_{\rho\sigma,\mu}-2
{}^\perp\!U_{\rho\mu,\sigma}\right) = 0\, .
\end{gather*}
This is the appropriate modification of (\ref{gunther_1}), the
novelty being that this condition is \emph{first derivative} in
$U$. Define $
 \Psi_{\mu}:= g^{\rho\sigma} \left({}^\perp\!U_{\rho\sigma,\mu}-2
{}^\perp\!U_{\rho\mu,\sigma}\right)$. The candidate Reduced
Einstein equations for our system will be
\begin{gather*}
 R_{\mu\nu} - \Psi_{(\mu,\nu)} = S_{\mu\nu}\, ,
\end{gather*}
together with initial value constraints $\Psi_{\mu}|_\Sigma = 0$,
$G_{0\mu}|_{\Sigma} = 0$. Now in order to get a hyperbolic system
we need to give some wave equation $\tilde\Box{}^\|\!U_\mu =
\cdots$. In the method of Gunther, the right hand side of
(\ref{gunther_1}) was fixed by the need to reduce
(\ref{starting_point}) to second order elliptic form
(\ref{gunther_2}). In our case, the tangential perturbations may
be given arbitrary dynamics and it is the isometric bending
degrees of freedom contained in the normal perturbations which
must be fixed in a specific way in order to obtain a hyperbolic
system. This difference is a consequence of the geometrical nature
of the Einstein equation. We shall see that our choice of
isometric bending gauge condition, namely $\Psi_\mu =0$, can be
implemented by initial value constraint which is conserved by the
evolution equations and that, for technical reasons, a preferred
canonical choice of wave equation for $\pau$ is indeed $\Gamma_\mu
-\Psi_\mu =0$.

\subsection{The nonlinear theory}

We now turn to consider (\ref{embedded_Einstein}) in more detail.
We recall that $\bX = \bhX +\bU$ and that both the covariant
derivatives and the inverse metric used to raise indices depend on
the physical metric. Therefore (\ref{embedded_Einstein}) is highly
nonlinear in $U$.

Introduce the noncovariant derivative operator $\tilde{\Box} :=
g^{\mu\nu}\partial_\mu\partial_\nu$. The Ricci tensor may be
re-expressed as:
\begin{gather}
 R_{\mu\nu} - \Gamma_{(\mu,\nu)} =  -\frac{1}{2}\tilde{\Box}h_{\mu\nu}+ \Gamma^\rho_{\sigma\nu}\Gamma^{\ \, \sigma}_{\rho\ \, \mu}
 -\Gamma_\rho \Gamma^\rho_{\mu\nu}
 +2 \Gamma^{\rho\,\sigma}_{\ \ (\mu}\Gamma_{\nu)\, \rho\sigma}\,
 .\label{non_lin_reduced_Einstein}
\end{gather}
Define:
\begin{gather}
 \Psi_{\sigma\, \mu\nu}: = - \peu_{\sigma\mu,\nu} - \peu_{\sigma\nu,\mu}
 + \peu_{\mu\nu,\sigma} \, ,
 \qquad
 \Psi_\sigma := g^{\mu\nu}\Psi_{\sigma\, \mu\nu} \, .
\end{gather}
Then we have the following useful identities for the christoffel
symbols:
\begin{align}
 \Gamma_{\sigma\, \mu\nu} & = \pau_{\sigma,\mu\nu}
  + \bU_{,\sigma}\cdot\bU_{,\mu\nu} +\Psi_{\sigma\, \mu\nu}\, ,
  \\
  \Gamma_\sigma & = \tilde\Box \pau_\sigma +
  \bU_{,\sigma}\cdot\tilde\Box \bU +\Psi_\sigma\, .
\end{align}
Applying $-\frac{1}{2}\tilde{\Box}$ to (\ref{starting_point}) we
have:
\begin{align}
 -\frac{1}{2} \tilde{\Box}h_{\mu\nu}
  & = \tilde{\Box} \peu_{\mu\nu} +
 \tilde{\Box}\bU\cdot\bU_{,\mu\nu} - \bU_{,\mu\rho}\cdot
 \bU_{,\nu\sigma}g^{\rho\sigma}\notag
 \\& \quad - \Gamma_{(\mu,\nu)} + \Psi_{(\mu,\nu)}
  -2\Gamma^{\rho\sigma}_{\ \  (\mu } \Gamma_{\nu)\rho\sigma}
  + 2 \Gamma^{\rho\sigma}_{\ \  (\mu} \Psi_{\nu)\rho\sigma}
  \label{non_lin_box_h}
\end{align}
Combining (\ref{non_lin_reduced_Einstein}) and
(\ref{non_lin_box_h}) we obtain:
\begin{align}
 R_{\mu\nu} -\Psi_{(\mu,\nu)}
 &= \tilde{\Box} \peu_{\mu\nu} +
 \tilde{\Box}\bU\cdot\bU_{,\mu\nu} - \bU_{,\mu\rho}\cdot
 \bU_{,\nu\sigma}g^{\rho\sigma}
  + 2 \Gamma^{\rho\sigma}_{\ \  (\mu} \Psi_{\nu)\rho\sigma}
  + \Gamma^\rho_{\sigma\nu}\Gamma^{\ \, \sigma}_{\rho\ \, \mu}
 -\Gamma_\rho \Gamma^\rho_{\mu\nu}\, . \label{appropriate_form}
\end{align}

We introduce the equation system:
\begin{align}
 R_{\mu\nu} -\Psi_{(\mu,\nu)} &= S_{\mu\nu}\, ,\label{R-Psi}
 \\
 \Gamma_\mu - \Psi_\mu &= 0\, ,
\end{align}
which, explicitly reads:
\begin{align}
 \tilde{\Box} \peu_{\mu\nu} & =
 - \tilde{\Box}\bU\cdot\bU_{,\mu\nu} + \bU_{,\mu\rho}\cdot
 \bU_{,\nu\sigma}g^{\rho\sigma}
 - 2 \Gamma^{\rho\sigma}_{\ \  (\mu} \Psi_{\nu)\rho\sigma}
  - \Gamma^\rho_{\sigma\nu}\Gamma^{\ \, \sigma}_{\rho\ \, \mu}
 + \Gamma_\rho \Gamma^\rho_{\mu\nu} + S_{\mu\nu}\label{explicit_1}
\\
 \tilde\Box \pau_\sigma & = -
  \bU_{,\sigma}\cdot\tilde\Box \bU\, ,\label{explicit_2}
\end{align}
and the initial value constraints:
\begin{align}
 \Psi_\mu|_\Sigma = 0\, , \label{Initial_Psi}
 \\
 n^\nu  (G_{\mu\nu}-T_{\mu\nu})|_\Sigma = 0\, ,\label{initial_R}
\end{align}
where $n^\mu$ is a time-like normal vector orthogonal to $\Sigma$.
By going to an orthonormal basis $e_{\bar\mu}$ such that
$e_{\bar{0}}$ is normal to $\Sigma$ we observe that
$G_{\bar{0}\bar{0}} = \sum_{\bar{i},\bar{j} =1}^3
(\bX_{;\bar{i}\bar{i}}\cdot \bX_{;\bar{j} \bar{j}} -
\bX_{;\bar{i}\bar{j}}\cdot\bX_{;\bar{j} \bar{i}})$ and
$G_{\bar{0}\bar{a}} = \sum_{\bar{i}=1}^3
(\bX_{;\bar{0}\bar{a}}\cdot \bX_{;\bar{i} \bar{i}} -
\bX_{;\bar{0}\bar{i}}\cdot\bX_{;\bar{a} \bar{i}})$ do not depend
on $\bX_{;\bar{0} \bar{0}}$. Substituting the initial value
constraints into (\ref{R-Psi}) we obtain
$n^\nu \Psi_{\mu,\nu} |_\Sigma =0$. Taking the covariant
divergence of (\ref{R-Psi}), using the Bianchi identity and
$T_{\mu\nu}^{\ \ ;\nu} =0$ we obtain
\begin{gather}
 \tilde{\Box}\Psi_\mu = 2 \Gamma^{\rho\sigma}_{\ \ \mu} \Psi_{(\rho,\sigma)}
 + 2 \Gamma^\nu \Psi_{(\mu,\nu)}\, .
\end{gather}
Therefore (\ref{Initial_Psi}) is conserved and any solution of our
system satisfies $G_{\mu\nu} = T_{\mu\nu}$, $\Gamma_\mu = \Psi_\mu
=0$.

Equations (\ref{explicit_1}) and (\ref{explicit_2}) constitute a
nonlinear hyperbolic equation system  for $U$. We note that the
right hand side of (\ref{explicit_2}) is somewhat arbitrary and
could instead be chosen to be zero. The choice we have made is so
as to recover $\Gamma^\mu =0$. Regardless of this choice the
equation system is far from being quasilinear. The nonlinear terms
in (\ref{explicit_1}) contain second derivatives of $U$ explicitly
and also implicitly in the $\Gamma \Gamma$ terms. We will now show
that, whilst the former seem to be an essential feature, the
latter can be effectively smuggled into the initial conditions.

\subsection{Reformulating the nonlinear theory by including an auxhiliary metric
field}

Let us introduce now a metric ${\frak g}$ as an auxiliary
variable, a priori independent of $U$. We first define
\begin{align}
 \Phi_{\mu\nu}&:= -2 \peu_{\mu\nu} + 2 \pau_{(\mu,\nu)}
 + \bU_{,\mu}\cdot \bU_{,\nu} +\eta_{\mu\nu} -{\frak g}_{\mu\nu}
  = g_{\mu\nu} - {\frak g}_{\mu\nu}\, ,
 \\
 \varPsi_\mu  &:= {\frak g}^{\rho\sigma} (-2 \peu_{\sigma\mu,\nu}
   + \peu_{\mu\nu,\sigma})
\end{align}
so that for an isometric embedding $\Phi_{\mu\nu} =0$ and
$\varPsi_\mu = \Psi_\mu$. Consider now the equation system:
\begin{align}
 {\frak g}^{\rho\sigma} \peu_{\mu\nu, \rho\sigma} & =
 - {\frak g}^{\rho\sigma}\bU_{,\rho\sigma}\cdot\bU_{,\mu\nu} + \bU_{,\mu\rho}\cdot
 \bU_{,\nu\sigma}{\frak g}^{\rho\sigma}
 - 2 \Gamma^{\rho\sigma}_{\ \  (\mu} \Psi_{\nu)\rho\sigma}
  - \Gamma^\rho_{\sigma\nu}\Gamma^{\ \, \sigma}_{\rho\ \, \mu}
 + \Gamma_\rho \Gamma^\rho_{\mu\nu} + S_{\mu\nu}\, ,\label{explicit_auxiliary_1}
\\
 {\frak g}^{\rho\sigma} \pau_{\mu,\rho\sigma} & = -
  \bU_{,\sigma}\cdot{\frak g}^{\rho\sigma}\bU_{,\rho\sigma}\, ,\label{explicit_auxiliary_2}
\\
 -\frac{1}{2} {\frak g}^{\rho\sigma} {\frak g}_{\mu\nu,\rho\sigma}  & =
 - \Gamma^\rho_{\sigma\nu}\Gamma^{\ \, \sigma}_{\rho\ \, \mu}
 +\Gamma_\rho \Gamma^\rho_{\mu\nu}
 -2 \Gamma^{\rho\,\sigma}_{\ \ (\mu}\Gamma_{\nu)\, \rho\sigma}
 + S_{\mu\nu}\, , \label{non_lin_hyp_1}
\end{align}
with initial conditions:
\begin{align}
 n^\nu  (G_{\mu\nu}-T_{\mu\nu})|_\Sigma & = 0\, ,
\\
  \varPsi_\mu|_\Sigma  & =0\, ,
\\
 \Phi_{\mu\nu}|_\Sigma &= 0\, ,
 \\
 n^\rho \Phi_{\mu\nu,\rho}|_\Sigma &= 0\, . \label{dangerous_initial_condition}
\end{align}
In
(\ref{explicit_auxiliary_1})-(\ref{dangerous_initial_condition})
we regard $\Gamma$ with indices in various positions and
$G_{\mu\nu}$ to be functions of ${\frak g}$ and its derivatives
rather than functions of $U$.

Equation (\ref{dangerous_initial_condition}) is not as it stands
an equation for initial data. But, going again to an adapted
frame, we find that the second time derivatives appear in the
combination $\pau_{\bar\mu|\bar{0}\bar{0} } +
\bU_{|\bar\mu}\cdot\bU_{|\bar{0}\bar{0}}$. Therefore we can use
evolution equation (\ref{explicit_auxiliary_2}) to eliminate them.

Taking the difference between (\ref{explicit_auxiliary_1}) and
(\ref{non_lin_hyp_1}) and using (\ref{explicit_auxiliary_2}) we
obtain:
\begin{align}
 {\frak g}^{\rho\sigma} \Phi_{\mu\nu,\rho\sigma}
  & =  \Gamma^{\rho\,\sigma}_{\ \ \,\mu}\left[
 \Phi_{\nu\rho,\sigma}
 + \Phi_{\nu\sigma,\rho}
 -\Phi_{\rho\sigma,\nu}\right]
  + \Gamma^{\rho\,\sigma}_{\ \ \,\nu}\left[
 \Phi_{\mu\rho,\sigma}
 + \Phi_{\mu\sigma,\rho}
 -\Phi_{\rho\sigma,\mu}\right]
  \, .
\end{align}
Therefore the constraint for $\Phi_{\mu\nu}$ is conserved, our
embedding is isometric, equations (\ref{non_lin_hyp_1}) and
(\ref{explicit_auxiliary_1}) become equivalent and our system
reduces to the previous one  (\ref{explicit_1}-\ref{initial_R}).
Conservation of the remaining constraints then follows and
solutions of the system satisfy:
\begin{gather}
R_{\mu\nu} = S_{\mu\nu}\, , \qquad \Gamma_\mu =\Psi_\mu = 0\, ,
\qquad {\frak g}_{\mu\nu} = g_{\mu\nu}\, .
\end{gather}

By introducing ${\frak g}_{\mu\nu}$ as an auxiliary field we
obtained a kind of hybrid between the standard formulation of GR
in De Donder gauge by Choquet-Bruhat, and a hyperbolic version of
Gunther's perturbation method. In this formulation, the nonlinear
term on the right hand side of equation
(\ref{explicit_auxiliary_2}) is not arbitrary. It is in just the
right form to allow us to conserve the initial condition $g =
{\frak g}$.
The evolution equations are of the form:
\begin{align}
  {\frak g}^{\rho\sigma}(\bhX_{,\mu\nu} \cdot \bU)_{,\rho\sigma} & =
 {\frak g}^{\rho\sigma}\left(
 \bU_{,\mu\nu}\cdot\bU_{,\rho\sigma}
 - \bU_{,\mu\rho}\cdot\bU_{,\nu\sigma}\right) +\cdots\, , \label{hyperbolic_2}
\\
  {\frak g}^{\rho\sigma} (\bhX_{,\mu} \cdot \bU)_{,\rho\sigma} & =
 -  {\frak g}^{\rho\sigma}\bU_{,\mu}\cdot\bU_{,\rho\sigma}\,
 ,\label{hyperbolic_3}
\\
  -\frac{1}{2} {\frak g}^{\rho\sigma}  {\frak g}_{\mu\nu,\rho\sigma}
  & = \cdots\, .\label{hyperbolic_4}
\end{align}
where the ellipsis denotes terms of at most first order in
derivative. Since $\bhX$ is free, (\ref{hyperbolic_2}) and
(\ref{hyperbolic_3}) can be expressed in the form:
\begin{align}\label{hyp_matrix_equation}
    {\frak g}^{\rho\sigma} \partial_\rho\partial_\sigma U^A =
    (\hat{M}^{-1})^A_{\ B}\,
    {\cal Q}^B( {\frak g},\partial {\frak g}; \bU, \partial\bU, \partial\partial \bU)  \, .
\end{align}
where $\hat{M} = (\partial \partial \hat{X},\partial \hat{X})$
understood as a 14 by 14 matrix. Second derivatives in ${\cal Q}$ occur only in terms which are quadratic order in $U$ or derivatives. ${\cal Q}$ also depends
on the background embedding up to its fourth derivatives.

Equations (\ref{hyperbolic_4}-\ref{hyp_matrix_equation})
constitute a second order fully non-linear differential equation
system which is hyperbolic about $U = 0$. Returning to the caveat mentioned at the beginning of the section, note that (\ref{hyp_matrix_equation}) is linear in $U_{,00}$ and can be algebraically solved for it, allowing us to get a bound on it in terms of the initial data. Therefore, for small enough initial data, the perturbed embedding is guaranteed to be free, at least for some period of time. To see that the evolution must be well defined for some finite time period, we note that it is possible to eliminate $U_{;00}$ from the right hand side of (\ref{hyp_matrix_equation}). Then, by introducing new variables equal to spatial derivatives of the original ones, one can construct a new system, which is quasilinear. The principal symbol will be modified, but as long as all fields remains small, the equations will remain hyperbolic. This will be discussed more fully in a follow up article. Another matter which deserves further study is the question of solving the initial data, which basically amounts to the perturbation problem for the initial surface subject to the isometric bending constraints $\Psi_\mu =0$.

\subsection{Comments on the linearised theory}

By considering the action for the linearised theory, we see
another explanation for why we obtained a wave equation by
imposing (\ref{Initial_Psi}). The massless Fierz-Pauli action is:
\begin{gather}\label{lin_EW}
 I^\text{(lin)} =  \frac{1}{4}\int_\mathcal{M}
 \left(2 h_{\mu\nu,\rho} h^{\mu\rho,\nu}
 - 2 h_{\mu\nu}^{\ \ , \nu} h_{\rho}^{\ \rho, \mu}
 + h_{\mu}^{\ \mu, \nu}h^{\rho}_{\ \rho,\nu}
 - h_{\mu\nu,\rho} h^{\mu\nu,\rho}\right)
\end{gather}
In terms of the embedding coordinates,
we have
\begin{align}
 I^\text{(lin)} & =  -\int_\mathcal{M}
 2 { {}^{\perp}\! U}_{\mu\nu,\rho} { {}^{\perp}\! U}^{\mu\rho,\nu}
 - 2 { {}^{\perp}\! U}_{\mu\nu}^{\ \ , \nu} { {}^{\perp}\! U}_{\rho}^{\ \rho, \mu}
 + { {}^{\perp}\! U}_{\mu}^{\ \mu, \nu}{ {}^{\perp}\! U}^{\rho}_{\ \rho,\nu}
 - { {}^{\perp}\! U}_{\mu\nu,\rho} { {}^{\perp}\! U}^{\mu\nu,\rho} +
  \partial_\mu \Omega^\mu
\end{align}
with
\begin{gather*}
 \Omega^\mu := 2 \pau^\mu_{\ ,\rho\sigma} \peu^{\rho\sigma}
 -2 \pau_{\rho, \ \sigma}^{\ \ \rho} \peu^{\sigma\mu}
 + \pau_{\rho ,}^{\ \ \mu\rho} \peu_{\sigma}^{\ \sigma}
 - \pau^{\mu\ \ \rho}_{\ , \rho}\peu_{\sigma}^{\ \sigma}
 + \pau^{\rho\sigma}\pau^{\mu}_{\ ,\rho\sigma}
 - \pau_{\rho , }^{\ \ \mu}\peu_{\sigma , }^{\ \ \sigma \rho}
\end{gather*}
Viewed from this point of view it is trivial to establish
equivalence with the metric theory - it is simply a matter of
replacing $h$ with ${}^\perp\!U$. However, the interpretation is
somewhat different. In the metric theory the gauge invariance
$h_{\mu\nu} \to h_{\mu\nu} + v_{\mu,\nu}+v_{\nu,\mu}$ is
associated with infinitesimal coordinate transformations. In the
embedding theory the automorphisms of the surface are associated
with $\pau_\mu \to \pau^{\prime}_\mu$ which is manifested in the
fact that $\pau$ drops out of the linearised action, whereas the
transformations $\peu_{\mu\nu} \to \peu_{\mu\nu} + v_{\mu,\nu} + v_{\nu,\mu}$ are
geometrically nontrivial from the extrinsic point of view. They
have the status of pure gauge only because the
action is an intrinsic invariant.
 In view of this,
we are at liberty to modify the action to promote the isometric
bending to physical degrees of freedom without breaking intrinsic
diffeomorphism invariance. It would be interesting to pursue some modification such as massive gravity via this route.


\section{Further discussion}

\subsection{Equivalence with GR at the linearised level?}

In ref. \cite{DeserPiraniRobinson76} it was pointed out that, for
typical examples of embeddings of interest, the degrees of freedom
of linearised GR tend to show up only at quadratic order in the
embedded theory. For example, consider the embedding of the
Schwarzschild space-time into 5+1 dimensions due to
Fronsdal\cite{Fronsdal}. A perturbation of the Fronsdal embedding
has some non-trivial content at the linearised level. Adding extra
dimensions (in straightforward fashion to create a product space)
does not add anything at linear level\cite{Kerner:2008nt}.
However, degrees of freedom are certainly missing. Why? As was
pointed out in section \ref{Free_section}, the linearised embedded
theory fails to encompass the linearised metric theory to the
extent to which the osculating space is degenerate. Adding trivial
extra dimensions will not remedy the problem. Instead we must find
an alternative embedding which is free. Now local free embeddings
are generic in high enough dimension. In $N \geq 14$ (or 10 for
analytic data) the set of embeddings for which the full linearised
theory does not show up is therefore expected to be of measure
zero. The problem is that this measure zero set consists of all
the concrete examples that are well known to Relativists. So one
must be careful of intuition based on specific examples until
appropriate examples are found.

Let us consider now how such examples may be constructed. One way
is to take our favourite known embedding of $\mathcal{M} \to
E^{R,1}$ and compose it with a free embedding of $E^{R-1,1}\to
E^{N-1,1}$ for some N. This has the advantage that we only ever
need to find one explicit FIE. The disadvantage is that this is
not very economical at all with dimensions. For example, for the
Fronsdal embedding, we have R=6 which will require N = 27 even for
a local FIE. The other approach is to look directly for a free
embedding in each case of interest. There does not seem to be any
examples in the literature, although some spatially free
embeddings were given in Ref. \cite{Bustamante05}.

\subsection{The 13+1 dimensional theory}\label{14D_discussion}

The study of (\ref{hyperbolic_4}-\ref{hyp_matrix_equation}) is
equivalent to the study of the Einstein equations, for an embedded
surface close to a given freely embedded background, as an initial
value problem. This approach is useful for identifying appropriate
gauge fixing and perhaps for identifying the degrees of freedom.
It also allows us to confirm the intuitive picture of
gravitational waves as ripples propagating along the world-sheet of
the embedded manifold. However, it may not be the most appropriate
method for addressing long time existence and uniqueness of solutions to the
evolution equations. To attack directly such highly nonlinear equations may be a
very hard problem (although it seems some results are known in
certain cases\cite{Hormander}). Alternatively we may abandon the
idea of a wave equation for $U$. Indeed the problem can be broken
down into two steps: first one evolves the initial data for the
intrinsic metric using the Einstein equation, then one embeds the
resulting manifold, according to some initial constraints on the
embedding. For the second step the complication arises because we
have introduced the lorentzian physical metric into
(\ref{hyperbolic_2}) and (\ref{hyperbolic_3}). It may be better to
adapt existing methods which in one way or another introduce an
auxiliary euclidean metric. If we are sufficiently close to the
background, then the second step is a 4 dimensional
perturbation problem similar to that discussed in chapter
\ref{perturbation_section}. This would give a combined
hyperbolic/elliptic system. A metric perturbation will be
realisable as a perturbation of $\hat{X}$ when some condition like
(\ref{smallness}) or (\ref{smallness_h}) is satisfied.
Heuristically
\begin{gather*}
 \| h \| \lesssim \| \hat{M}^{-1}\|^{-2}\, .
\end{gather*}
Since the Einstein system is well posed, by making the initial
data small enough, we may ensure that the condition is met up to
arbitrarily large time. To ensure that, with  finite initial data,
it is satisfied for all time would require some strong stability result for the background spacetime.

More generally we might hope for some version of cosmic
censorship, whereby an initially free embedding will generically
remain: 1) an immersion; 2) free, except perhaps inside some
horizons. We might also add a stronger condition:  1') embedded.
Roughly speaking 1) is a poor mans cosmic censorship which says
generically there will be no naked singularities where the metric
becomes degenerate or singular\footnote{This is only roughly speaking because there are certain types of purely extrinsic sigularity that can occur in higher dimensions whereby the map $dX$ is not injective but the metric $dX \cdot dX$ is well defined e.g. the ``embedding" of a flat two-plane into Minkowski space $(x,y) \to (r,x,y,r)$.}; 2) is an additional
requirement to allow us to develope the embedding; 1') says that
$X(M)$ does not self-intersect. These are properties we would like
solutions to have, but at the moment we have no arguments in
favour of such a conjecture. It is not even easy to make a precise statement of it. Any statement in terms of inextendibility of developements is problematic. For example, even if an embedding can be developed out to infinity of $\mathbb{E}^{13,1}$, it is possible that as an intrinsic manifold, $M$ is extendible\cite{Poznyak-Sokolov}. This can happen if $X(M)$ asymptotically approaches a null surface of $\mathbb{E}$.

\subsection{Global embeddings. $N>14$?}
\label{Large_N_section}

In order to consider the global problem, and accomodate a general
background, we will presumably need to go to higher than 13+1
dimensions. The current lower bound on $N$ for embedding any
Riemannian $n$-manifold is $N_* = \min (\frac{1}{2}n (n+5),
\frac{1}{2}n(n+3) +5)$\cite{Gunther90}. For globally hyperbolic
Lorentzian manifolds it is $N_* +1$ \cite{Mueller:2008gc}. The way
in which $N_*$ is obtained is a long story which we are not able
to summarise here, so we will simply give some
comments. The $\frac{1}{2}n(n+5)$ comes purely from differential
topology- basically the generalisation of Whitney's theorem to
free embeddings. Since we are restricting our interests to
globally hyperbolic manifolds diffeomorphic to $\mathbb{R}\times
\Sigma$, this number might conceivably be reduced. The
$\frac{1}{2}n(n+3)+5$ comes from requiring a 5 dimensional
orthogonal compliment to the osculating space with which to make
coarser adjustments to the induced metric without loosing control
of $M^{-1}$, so as to apply the perturbation result. Depending
upon whether this last number can be improved upon, we may need up
to 19+1 dimensions for our embedding space. Even if
$\mathcal{M}_4$ is flat, it is not clear whether a global free
embedding into 13+1 dimensions exists.

Let $\hat{X}:\mathcal{M} \to \mathbb{E}^{N-1,1}$ be free.
Introduce an orthonormal basis $\bm{e}_{I}$, $I = 15,\dots, N$, of
the orthogonal compliment to the osculating space. Then a
perturbation $U$ can be decomposed as
\begin{gather}
 \bm{U} =  \peu_{\mu\nu} (P^{-1})^{\mu\nu\rho\sigma}
 \bhX_{,\rho\sigma} + \pau_{\mu} g^{\mu\nu} \bhX_{,\nu}
 + Y^I \bm{e}_I \, .
\end{gather}
At the linearised level $Y$ will drop out of the Einstein
equations. At the nonlinear level
(\ref{hyperbolic_2}-\ref{hyperbolic_4}) will still apply, so that
$Y$ will contribute to the nonlinear self interaction terms, but
will not have any propagator. In order to recover a hyperbolic
system we may then proceed to treat $Y$ in the same way as $\pau$,
by giving some dynamics
\begin{gather}\label{proposed_Y_equation}
 \tilde\Box Y^I = \cdots
\end{gather}
which can presumably be interpreted as a gauge fixing. Certainly,
locally $Y$, like $\pau$, can be gauged away. But over long
timescales, $U$ may grow too large for the perturbation result to
apply, in which case $Y$ may need to be nonzero in order for a
solution to exist. In a sense, a global change in $Y$ is more like
changing the background rather than a component of the
perturbation. At the present, it is not clear whether a full global description of the gravitational field can be recovered in this way and if so what the appropriate form of (\ref{proposed_Y_equation}) should be.





\acknowledgements I thank J. Costa and A. Flachi for useful
discussions. I also thank S. Paston for a very fruitful
correspondence and for pointing out an important correction to the
first draft of this paper. This work was supported by the
Funda\c{c}\~{a}o para a Ci\^{e}ncia e a Tecnologia of Portugal and
the Marie Curie Action COFUND of the European Union Seventh
Framework Programme (grant agreement PCOFUND-GA-2009-246542).

\begin{appendix}

\section{Weighted Sobolev spaces}\label{sobolev_appendix}

Let $e$ be the Euclidean metric on $\mathbb{R}^3$ and $\mu_e$ be
the standard volume element. We shall be interested in quantities
which decay faster than $r^{\delta }$ so let us introduce the
quantity $\sigma := \sqrt{1+r^2}$ which is asymptotically $\sigma
\sim r$ but has the advantage of being positive definite.

\begin{definition}
$H_{s,\delta} = W^2_{s, \delta-3/2}$, $s\in \mathbb{N}$, $\delta
\in \mathbb{R}$ is the space of functions or tensor fields $f$
over $\mathbb{R}^n$ with square-integrable weak derivatives of
order up to $s$ such that
\begin{gather}
 \| f \|_{s,\delta} :=\left( \sum_{\alpha\leq s}
 \int_{\mathbb{R}^3}
 \sigma^{2(|\alpha| +\delta -3/2) } |\partial^\alpha f|^2 \mu_e\right)^{1/2} < \infty
\end{gather}
where $| f |$ denotes the pointwise norm of tensor $f$ in the
metric $e$.
\end{definition}
We shall need the following elementary properties which follow
directly from the definition: $\| K f\|_{s,\delta} = |K|
\|f\|_{s,\delta}$ for $K$ constant; $\|f_{,a}\|_{s-1,\delta+1}
\leq \|f\|_{s,\delta}$; $
 \| \sum_c f_{,ac}  f_{,bc}\| \leq C_1 \| f_{,ab}\| \| f_{,cd}\|$;
$ \| \Delta f \| \leq C_2 \| f_{,ab}\|$ for some constants $C_1$,
$C_2$. We shall also need the following two well known properties:
\begin{theorem}[Continuous multiplication rule]\label{sobolev_product} If $ s_1, s_2 \geq s\, , \  s_1 +s_2 > s +3/2\, , \ \delta_1 +\delta_2 >
  \delta$ then  $ f \in  H_{s_1,\delta_1}\, ,\  g \in H_{s_2 ,\delta_2 }\
  \Rightarrow \
  f \otimes g \in H_{s,\delta}$ and, for some constant $C$:
\begin{gather*}
 \|f\  g\|_{s,\delta} \leq C \|f\|_{s_1,\delta_1}\|g\|_{s_2,\delta_2} \, .
  \end{gather*}
\end{theorem}

\begin{theorem}[Invertibility and open mapping property of
Laplacian]\label{invert_Laplacian} If  $s\geq 2$, $0<\delta<1$
then
 $\Delta: H_{s,\delta} \to
 H_{s-2,\delta +2}$ is an isomorphism and for some constant $C$:
\begin{gather*}
 \|f\|_{s,\delta} \leq C \|\Delta f\|_{s-2,\delta+2} \, .
  \end{gather*}
\end{theorem}

\section{Proof of theorem \ref{Perturbation_Sobolev}}\label{proof_appendix}

Let  $\hat{X}$ be a FIE of $(\mathbb{R}^3, \gamma)$ into
$\mathbb{E}^N$ satisfying $M^{-1}  = K + L$ where $K$ is a
constant matrix. Assume and $L \in H_{q,\beta}$ and  $M^{-1}H \in
H_{p,\alpha}$ . We prove the existence of a FIE $X$ for
$(\mathbb{R}^3,\gamma + h)$ under the conditions of theorem
\ref{Perturbation_Sobolev} .

Let $U = \lambda W$ where
\begin{gather}
 \lambda :=  \sqrt{\frac{\| M^{-1} H \|_{p,\alpha} }{|K| + \| L\|_{q,\beta} }
 }
 \leq D \sqrt{\| h\|_{p,\alpha} }\, .
\end{gather}
Then (\ref{gunther_matrix}) becomes
\begin{gather}
 W = {\cal O}(W)\, ,
 \\
 {\cal O}(W) := \frac{1}{\lambda} M^{-1}H + \lambda M^{-1}
 \Delta^{-1} Q(W)\, .
\end{gather}
Assume $W_m \in H_{p,\alpha}$ and consider the iteration $W_{m+1}
= {\cal O}(W_m)$ with $W_0 = \frac{1}{\lambda} M^{-1}H$ and prove
convergence to a solution by the contraction mapping theorem.
Suppose $W \in {\cal F} := \{ V\in H_{p,\alpha}\ ;\  \|
V\|_{p,\alpha} \leq 1\}$. Then
\begin{align}
 \|{\cal O}(W) \|_{p,\alpha} & \leq \frac{1}{\lambda}
 \| M^{-1}H\|_{p,\alpha} + \lambda |K|\, \| \Delta^{-1}
 Q(W)\|_{p,\alpha} + \lambda  \| L \Delta^{-1}
 Q(W)\|_{p,\alpha}
 \\
 & \leq \sqrt{\| M^{-1}H\|_{p,\alpha} (|K| + \|L\|_{q,\beta} )}
 \left(  1 + C_1 \| \Delta^{-1} Q(W)\|_{p,\alpha}\right)\, ,
 \label{L Deltainv_Q}
\\
 & \leq \sqrt{\| M^{-1}H\|_{p,\alpha} (|K| + \|L\|_{q,\beta} )}
 \left(  1 + C_2 \| Q(W)\|_{p-2,\alpha+2}\right)\,
 ,\label{Deltainv_Q}
\\
 & \leq \sqrt{\| M^{-1}H\|_{p,\alpha} (|K| + \|L\|_{q,\beta} )}
 \left(  1 + C_3 \|W^{\prime\prime}\|_{p-2,\alpha+2} \left(
 \|W^{\prime\prime}\|_{p-2,\alpha+2} + \|W^{\prime}\|_{p-1,\alpha+1}\right)
 \right) \, ,\label{expand_Q}
\\
 & \leq \sqrt{\| M^{-1}H\|_{p,\alpha} (|K| + \|L\|_{q,\beta} )}
 \left(  1 + C_4 \|W\|_{p,\alpha}^2\right)\, ,
\\
 & \leq C_5 \sqrt{\| M^{-1}H\|_{p,\alpha} (|K| + \|L\|_{q,\beta} )} \, .
\end{align}
The above inequalities are applications of Theorems
\ref{sobolev_product} and \ref{invert_Laplacian}, which are valid
provided we require certain conditions in $p, q, \alpha,\beta$:
(\ref{expand_Q}) requires $p\geq 4$; (\ref{L Deltainv_Q}) requires
$\alpha, \beta
>0$, $q\geq p$. These are sufficient to ensure all other
inequalities hold.

 Therefore ${\cal O}: {\cal F}\to {\cal F}$ provided that:
\begin{gather}
 \| M^{-1}H\|_{p,\alpha} (|K| + \|L\|_{q,\beta} ) \leq \frac{1}{C_5^2}\, .
\end{gather}
Now consider $W_1,\, W_2 \in {\cal F}$.
\begin{align}
 \|{\cal O}(W_2) -{\cal O}(W_1) \|_{p,\alpha} & = \lambda \| M^{-1}
 \Delta^{-1} Q(W_2) -  M^{-1}
 \Delta^{-1} Q(W_1) \|_{p,\alpha}\\
 & \leq
  \lambda C_6 (|K| + \|L\|_{q,\beta} ) \|
 Q(W_2) -Q(W_1)\|_{p-2,\alpha+2}\, ,
\\
 & \leq \lambda C_7 (|K| + \|L\|_{q,\beta} )
 \Big(\| W_2^{\prime\prime} + W_1^{\prime\prime}\|_{p-2,\alpha+2}
 \| W_2^{\prime\prime} -
 W_1^{\prime\prime}\|_{p-2,\alpha+2}\Big.\notag
\\
 & \qquad\qquad\qquad\qquad\qquad
 + \| W_2^{\prime} + W_1^{\prime}\|_{p-1,\alpha+1}
 \| W_2^{\prime\prime} - W_1^{\prime\prime}\|_{p-2,\alpha+2}
\\
 & \qquad\qquad\qquad\qquad\qquad \Big.
 +\| W_2^{\prime\prime} + W_1^{\prime\prime}\|_{p-2,\alpha+2}
 \| W_2^{\prime} - W_1^{\prime}\|_{p-1,\alpha+1}\Big)
 \notag
\\
  & \leq \lambda C_8 (|K| + \|L\|_{q,\beta} )
  \| W_2 + W_1\|_{p,\alpha}\| W_2 - W_1\|_{p,\alpha}
  \\
   & \leq C_9 \sqrt{\| M^{-1}H\|_{p,\alpha} (|K| + \|L\|_{q,\beta} )} \| W_2 - W_1\|_{p,\alpha}
\end{align}
Therefore ${\cal O}:{\cal F}\to {\cal F} $ is a contraction
mapping if
\begin{gather}
 \| M^{-1}H\|_{p,\alpha} (|K| + \|L\|_{q,\beta} ) < \max\left\{\frac{1}{C_5^2}, \frac{1}{C_9^2}\right\}\, .
\end{gather}
Applying the contraction mapping theorem we obtain the desired
result. \qed

\end{appendix}


\begin{thebibliography}{99}


\bibitem{Nash56}
J. Nash,
Ann. Math. {\bf 63} 20-63 (1956).

\bibitem{Greene70}
R. Greene,
Mem. Amer. Math. Soc., No. 97, Amer. Math. soc.,
Providence, RI, 1970.

\bibitem{Clarke70}
C. J. S. Clarke,
Proceedings of the Royal Society of
London, {\bf A} 314 417-428 (1970).

\bibitem{Gromov_Rokhlin}
M. L. Gromov and  V. A. Rokhlin,
Usp. Mat. Nauk 25,3-62 (Russian); Russ. Math.
Surv. 25 (1970) 1 -57.

\bibitem{Mueller:2008gc}
  O.~Mueller and M.~Sanchez,
  Trans.\ Am.\ Math.\ Soc.\  {\bf 363}, 5367 (2011)
  [arXiv:0812.4439 [math.DG]].



\bibitem{ReggeTeitelboim75}
  T.~Regge and C.~Teitelboim,
  In *Trieste 1975, Proceedings, Marcel Grossmann Meeting
  On General Relativity*, Oxford 1977, 77-87

\bibitem{PastonFranke07}
  S.~A.~Paston and V.~A.~Franke,
  Theor.\ Math.\ Phys.\  {\bf 153}, 1581 (2007)
  [arXiv:0711.0576 [gr-qc]].

\bibitem{PastonSemenova10}
  S.~A.~Paston and A.~N.~Semenova,
  Int.\ J.\ Theor.\ Phys.\  {\bf 49}, 2648 (2010)
  [arXiv:1003.0172 [gr-qc]].

\bibitem{Faddeev09}
  L.~D.~Faddeev,
  arXiv:0911.0282 [hep-th].

\bibitem{Pavsic-1986}
M Pavsic, Class. Quantum Grav. {\bf 2} 869 (1985).

\bibitem{Paston:2011db}
  S.~A.~Paston,
  Theor.\ Math.\ Phys.\  {\bf 169}, 1611 (2011)
  [arXiv:1111.1104 [gr-qc]].


\bibitem{Joseph:1962zz}
  D.~W.~Joseph,
  Phys.\ Rev.\  {\bf 126}, 319 (1962);
  Rev. Mod. Phys. {\bf 37}, 225 (1965).

\bibitem{DeserPiraniRobinson76}
  S.~Deser, F.~A.~E.~Pirani and D.~C.~Robinson,
  Phys.\ Rev.\ D {\bf 14}, 3301 (1976).


\bibitem{Bustamante05}
  M.~D.~Bustamante, F.~Debbasch and M.~-E.~Brachet,
  gr-qc/0509090.


\bibitem{Friedman_Review}
A. Friedman,
 Rev. Modern Phys. 37 (1965), 201-203.




\bibitem{Andrews}
B. Andrews, ``Notes on the isometric embedding problem and the
nash-moser implicit function theorem," Surveys in analysis and
operator theory (Canberra, 2001), 157-208, Proc. Centre Math.
Appl. Austral. Nat. Univ., 40, Austral. Nat. Univ., Canberra,
2002.

\bibitem{HanHong06}
Q. Han and J. Hong, ``Isometric Embedding of Riemannian Manifolds
in Euclidean Spaces,'' Mathematical surveys and monographs, AMS
(2006).



\bibitem{Private_comm_Paston}
S. Paston, Private communication.


\bibitem{Jacobowitz74}
 H. Jacobowitz,
 J. Diff. Geo. {\bf 9}, 291-307 (1974).



\bibitem{Friedmann61}
A. Friedman,
J. Math. Mech. {\bf 10} (1961)
625-650.




\bibitem{Choquet-Bruhat09}
Y. Choquet-Bruhat, ``General Relativity and the Einstein
Equations," Oxford (2009).


\bibitem{Anderson:2004zu}
  E.~Anderson,
  gr-qc/0409122.


\bibitem{Gunther89}
M. Gunther,
Annals of Global analysis and
Geometry {\bf 7} (1989) 69-77.

\bibitem{Gunther90}
M. Gunther, Zum Einbettungssatz von J. Nash. Math. Nachr. 144
(1989) 165-187; ``Isometric Embeddings of Riemannian Manifolds",
in Proceedings of the International Congress of Mathematicians,
Kyoto 1990 (Vol. 2), Springer-Verlag, New York (1991) pp
1137-1143.

\bibitem{OMurchada86}
N. O Murchada,
J. Math. Phys. {\bf 27} 2111 (1986); R. Bartnick, Comm.Pure Appl.
Math. {\bf 39} 661-693 (1986)

\bibitem{CB-Christodoulou81}
Y. Choquet-Bruhat and D. Christodoulou, Acta Math. {\bf 146}, 129
(1981).


\bibitem{Hinterbichler11}
  K.~Hinterbichler,
  Rev.\ Mod.\ Phys.\  {\bf 84}, 671 (2012)
  [arXiv:1105.3735 [hep-th]].


\bibitem{Fronsdal}
C. Fronsdal, Physical Review, 116 (3), p. 778 (1959)

\bibitem{Kerner:2008nt}
  R.~Kerner and S.~Vitale,
  arXiv:0801.4868 [gr-qc].

\bibitem{Hormander}
L. Hormander,
 Bol. Soc. Brasil. Mat {\bf 20} 1 (1989) 1-27;
Microlocal Analysis and Nonlinear Waves, The IMA Volumes in
Mathematics and its Applications {\bf 30} (1991) 51-81.


\bibitem{Poznyak-Sokolov}
E. Poznyak and D. Sokolov,
Journal of Soviet Mathematics
{bf 14}, 1407 (1980).




\end{thebibliography}
 \end{document}